\documentclass{article}
\usepackage{spconf,amsmath,graphicx}
\usepackage{cite}
\usepackage{xcolor, soul}
\usepackage{amsmath}
\usepackage{amssymb}
\usepackage{graphicx}
\usepackage{enumitem}
\usepackage{todonotes}
\usepackage{caption}
\usepackage{subcaption}
\usepackage{hyperref}
\sethlcolor{yellow}
\usepackage{array}

\usepackage{xparse}
\NewDocumentCommand\mm{g}{%
  \IfNoValueF{#1}{{\color{blue} \textbf{(MM: #1)}}}%
  \IfNoValueT{#1}{{\color{blue} \textbf{(MM)}}}%
}

\NewDocumentCommand\Amir{g}{%
  \IfNoValueF{#1}{{\color{red} \textbf{(Amir: #1)}}}%
  \IfNoValueT{#1}{{\color{red} \textbf{(Amir)}}}%
}

\newcommand{\subhead}[1]{\noindent{\textbf{#1.}}}

\newcolumntype{P}[1]{>{\centering\arraybackslash}p{#1}}
\newcommand{\rr}{\mathop{{\rm I}\mskip-4.0mu{\rm R}}\nolimits}
\title{On securing cloud-hosted cyber-physical systems using trusted execution environments}  

\name{Amir Mohammad Naseri, Walter Lucia, Mohammad Mannan, Amr Youssef}
\address{Concordia Institute for Information Systems Engineering (CIISE)\\
Concordia Univerity, Montreal, Canada.}
\begin{document}
	%\ninept
	%
	\maketitle
%-------------------------------------------------------------------------------------%
\begin{abstract}
		\label{sec:abs}
Recently, cloud control systems have gained increasing attention from the research community as a solution to implement
% in implementing \mm{This implies that cloud-based control services are used widely - but my understading is that there are only academic proposals/ {\color{red} Amir: I think we can use "attention" instead of "popularity."}} 
networked cyber-physical systems (CPSs). Such an architecture can reduce deployment and maintenance costs albeit at the expense of additional security and privacy concerns.  
In this paper, first, we discuss state-of-the-art security solutions for cloud control systems, and their limitations. Then, we propose a novel control architecture based on  Trusted Execution Environments (TEE). We show that such an approach can potentially address major security and privacy issues for cloud hosted control systems. Finally, we present an implementation setup based on Intel Software Guard Extensions (SGX), and validate its effectiveness on a testbed system. 
% The obtained results confirm that the secure trusted controller implementation introduces a negligible computational overhead, making such a solution practical in CPSs with strict real-time requirements.

% Previously, various methods have been proposed to provide controller security in cloud, which are either vulnerable to attackers or not efficient enough, and in some cases, to implement them, the practical aspects of the system need to be re-examined. In this paper, with using trusted execution environment of Intel, Intel SGX, a method is proposed which, provide both security and privacy of the controller as well as communication channels. This model is efficient enough to be implemented in systems that do not have much computing power. Finally, the performance of the proposed model for a quadruple-tank process system is tested and reviewed, and the results show the efficiency and good performance of the model.
\end{abstract}
%------------------------------------------------------------------------------------%

\section{Introduction}
\label{sec:intro}
With the development of cloud services, the implementation of industrial control systems into the cloud has received increasing attention.
% . \mm{Same comment as in the abstract - popular? /{\color{red} Amir: I think we can use "attention" instead of "popularity."}}
%Several cloud services are available in the market providing high computing power. 
The use of such services saves on the cost of setting up and maintaining industrial control systems (ICS), as well as off-loading computationally expensive tasks. 
Moreover, when ICS are geographically distributed, these cloud services are highly available and accessible from different locations~\cite{mahmoud2019networked}. The main concern when using cloud services in such applications is the security and privacy of the cloud environment, and communication channels between the plant and the controller.   
% %
% % \todo[inline, color=green]{Are there any attacks reported for cloud-hosted controller? If yes, we need to report them here}
% \todo[inline, color=orange]{I cannot find any reported attack which directly aim the cloud-hosted controller of an industrial plant. need more search. \href{https://www.cloudfoundry.org/blog/cve-2017-8037/}{LINK1},
% \href{https://www.apriorit.com/dev-blog/523-cloud-computing-cyber-attacks}{LINK2},
% \href{https://threatpost.com/healthcare-2021-cyberattacks-covid-19-patient-data/161776/}{LINK3},
% }

% Over the years, numerous cyber-attacks targeting safety-critical industrial control systems have been reported. Examples of such attacks include the security incident at a Maroochy's water services in Queensland, Australia \cite{slay2007lessons}, the cyber attack on the Ukrainian power grid \cite{case2016analysis}, and the attack on a water system in Florida \cite{robles_perlroth_2021}. Different solutions have been proposed to preserve either the security or the privacy of cloud-based networked control systems, see e.g., \cite{pang2011secure,kogiso2015cyber}. However, as discussed in Section~\ref{sec:existing-solutions}, such approaches present some limitations, motivating the solution proposed in this paper.\\
Different approaches have been proposed to enhance the security and privacy of networked CPSs
% \mm{Which term to use? CPS or ICS? /{\color{red} Amir: I think it is better to use the term "CPS" because it is more general. I am not sure about that the the term "CPS" works for our figures or not.}} 
where the controller is hosted in a cloud infrastructure. For example, Zhou et al.~\cite{zhou2013achieving} propose the use of conventional cryptographic algorithms to secure  plant-to-cloud communication. 
% The use of conventional cryptographic algorithm , e.g. RSA and Elgamal encryption system, for encrypting just transmitted messages \cite{zhou2013achieving} was one of the first models proposed for this purpose. However, due to the fact that the controller is not encrypted in this method, this model is vulnerable from the controller inside the cloud service.
Kogiso and Fujita~\cite{kogiso2015cyber} propose the use of  homomorphic encryption to ensure that the controller's operations can be performed without decrypting the received data, and hence addressing the confidentiality problem in the cloud (in addition to securing communication channels). Homomorphic encryption-based solutions have received increasing attention by the CPSs community; for full homomorphic and Pailier's homomorphic based solutions, see e.g.,~\cite{kim2016encrypting,tran2020implementing, murguia2020secure,lin2018secure}. %and references therein.
However, these homomorphic solutions suffer from unavoidable limitations related to the arithmetic operations allowed by the homomorphic schemes, ciphertext size explosion, and computation overhead. For solutions targeting only securing communication channels cannot protect controller logic and data against a malicious or compromised cloud provider. %Moreover, no cryptographic solutions can be used to defend cloud-based CPSs against software integrity attacks performed against the controller deployed in the cloud.
For data and execution security in the context of IoT and CPS applications, Shepherd et al.~\cite{shepherd2016secure} survey and compare several existing secure and trusted computing environments such as Trusted platform Module (TPM), Secure Elements (SE), Trusted Execution Environments (TEEs), and Encrypted Execution Environments (E3). %However, to the best of the author's knowledge such technologies has not been used yet to secure cloud-hosted CPSs. 
% \begin{figure} [htbp]
%             \centerline{\includegraphics[width=0.6\columnwidth]{Figures/cloud_based_control_system.eps}}
%             \caption{Cloud-based networked control system setup}
%             \label{fig:cloud-based-setup}
%     \end{figure}

In this paper, we explore the use of encryption and trusted execution environments to secure plant-to-cloud communication channels and protect data and controller logic for cloud-hosted CPS applications. To understand performance implications of our approach, we also design and implement a simple prototype for the  quadruple  tank system~\cite{johansson2000quadruple}, using Intel SGX as our TEE. Our results indicate that the introduced overhead is negligible, and highly-scalable yet secure CPS applications can be designed for a cloud-deployment scenario. We hope that our initial results may be useful to the CPS security community and encourage the design of more efficient and secure TEE-based solutions compared to current schemes that rely mostly on conventional cryptographic mechanisms and homomorphic schemes. 
\section{System Setup and Threat Model}\label{sec: ncss}

%As illustrated in Fig.~\ref{fig:cloud-based-setup}, 
A  typical cloud-based, networked control system consists of following main components: the plant, the controller, the cloud, and the communication channels.  The \textit{plant} is the physical entity that we want to control. It is usually equipped with a set of \textit{actuators} 
%(converting the control inputs into physical actions) 
and \textit{sensors}. 
%(capable of measuring some variables characterizing the plant's current state). 
The \textit{controller} collects the sensor measurements and computes, according to a pre-defined control logic,  the control commands sent to the actuators.
% \mm{This sentence is confusing}. 
In a cloud-based networked setup, the controller and the plant are spatially distributed, and the controller logic is implemented in a \textit{cloud service} provider. The communication channels are used for a real-time and bi-directional exchange of data (e.g., sensor measurements and control inputs) between the plant and the controller.  
    % $\bullet$ \textbf{Plant} - The plant is the physical entity that we want to control. It is usually equipped with a set of actuators (capable of imposing inputs to the plant) and sensors (capable of measuring some variables describing the plant's current state).\\
    % $\bullet$ \textbf{Controller and Cloud} - The controller collects the sensor measurements and computes, according to a pre-define control logic,  the control commands used by actuators.
    % In a cloud-based networked setup, the controller and the plant are spatially distributed, and the controller logic is implemented in a cloud service provider.\\
    % $\bullet$ \textbf{Communication channels} - The communication channels are used for a real-time and bi-directional exchange of data (e.g. sensor measurements and control inputs) between the plant and the controller. 

%\subsection{Threat Model}\label{sec:threatmodel}
\subhead{Threat Model} We consider the following attacks that can affect  the privacy/security of the cloud-based CPS controllers.
%\begin{itemize} [leftmargin=*]
    %\item[-] 
    
    \textit{Attacks against the communication channels} - By adopting the conventional Dolev-Yao threat model~\cite{dolev1983security},  a malicious entity with access to the public communication channels is assumed to be able to eavesdrop on the transmitted data 
    % (sensor measurement and control inputs) 
    and/or modify their content. Therefore, potentially, the confidentiality and the integrity of the control system could be compromised. Indeed, such attackers can exploit the eavesdropped data to gain further information about the controlled system's behaviour and use their disruptive capabilities to launch sophisticated undetectable attacks such as replay, covert, zero-dynamics attacks~\cite{dibaji2019systems,teixeira2015secure}. %\vspace{-0.25cm}
    %\item[-] 
    
    \textit{Attacks against the cloud service} - If the cloud operator is malicious, or if the service is vulnerable, then an unauthorized entity (e.g., malware authors) might be able to gain access to the data transmitted between the plant and the controller, even if encrypted and authenticated communications are used. Indeed, such attackers could read the encryption key (key-management problem), intercept the transmitted data after decryption, and change the control logic (with the consequence of jeopardizing the whole control loop).
%\end{itemize}

%***************************************************************************%
  
\section{Existing Solutions}\label{sec:existing-solutions}
\begin{figure}
     \centering
     \begin{subfigure}[b]{0.48\columnwidth}
         \centering
         \includegraphics[width=0.99\columnwidth]{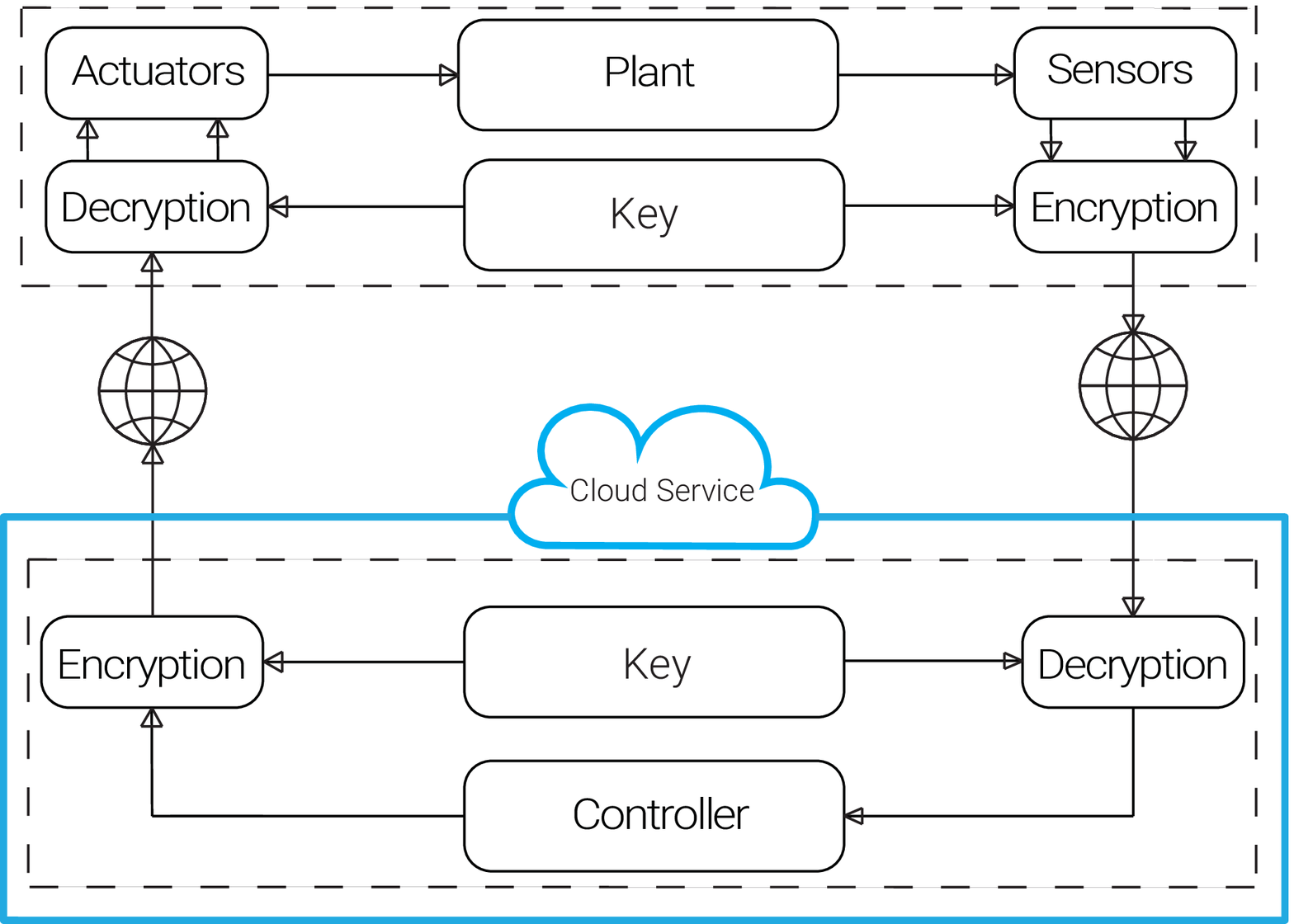}
         \caption{Encrypted communications.}
         \label{fig:solution-enchrypted-communications}
     \end{subfigure}
     \hfill
     \begin{subfigure}[b]{0.48\columnwidth}
         \centering
         \includegraphics[width=0.99\columnwidth]{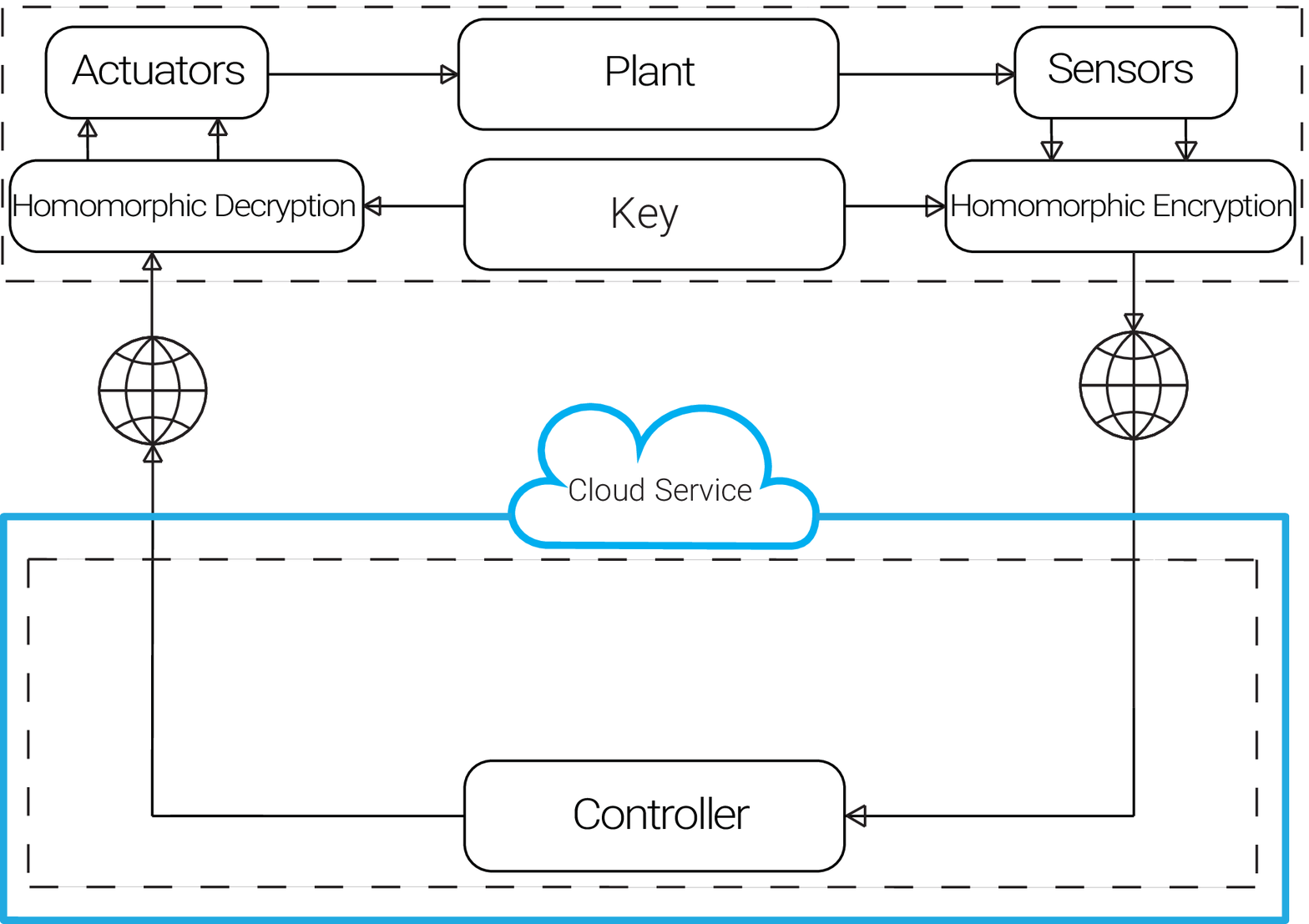}
         \caption{Homomorphic encryption.}
         \label{fig:solution-homomorphic-enchryption}
     \end{subfigure}
     \hfill
     \vspace{-0.3cm}
    \caption{Existing security solutions for cloud-based CPSs.} 
     \label{fig:existing-solutions}
    \end{figure}

\vspace{-0.3cm}
Different schemes have been proposed to secure networked control systems. A common solution is to use encrypted authenticated communications between the plant and the controller~\cite{patel2009improving}; see Fig.~\ref{fig:solution-enchrypted-communications}. Such a solution, at the cost of increased computational power to perform encryption/decryption operations at both the plant and controller's sides of the CPS, can mitigate the privacy and security issues related to cyber-attacks against the communication infrastructure. On the other hand, it does not address the security and privacy risks associated with the controller's deployment inside the cloud. %infrastructure. 

The use of homomorphic encryption has also been proposed to secure CPS solutions~\cite{kogiso2015cyber,farokhi2016secure}; see Fig~\ref{fig:solution-homomorphic-enchryption}. A distinctive capability of such a solution 
% (with respect to a generic encryption method) 
is that it allows the controller to implement the control logic (in terms of a additions and multiplications operations) directly on the received encrypted sensor measurements.
% , without decrypting them. 
Consequently, such an approach has the advantage of securing the communications while solving the privacy issues associated with the cloud infrastructure.
However, common drawbacks of homomorphic encryption include: the mathematical operations performed on the encrypted data are typically limited %(i.e., not all arithmetic operations are possible), 
and computationally expensive; and the plaintext to ciphertext bit expansion factor is usually very high.
Consequently, homomorphic-based solutions might not be practical for securing industrial control systems with fast sampling rate or narrow bandwidth. 

There are three different types of homomorphic encryption schemes, namely partially homomorphic encryption (PHE), somewhat homomorphic encryption (SHE), and fully homomorphic encryption (FHE). Each subclass is characterized by the set and number of encrypted operations allowed. Therefore, according to the limitations imposed by the used scheme, it might be challenging to recast any existing control algorithm into its encrypted counterpart. 
%
% Using different types of homomorphic encryption imposes constraints on the system, such as restrictions on arithmetic operations that can be applied to encrypted data, which can affects the performance of system. For these reasons, it is necessary to conduct detailed studies on different systems to use different types of homomorphic encryption to enhance both privacy and security of the system and in some cases it is necessary to redesign the controller.
For example, FHE allows an unlimited number of encrypted addition and multiplication operations and therefore it is particularly appealing to implement sophisticated control solutions such as dynamic feedback control or model predictive control. However, such a freedom comes with a  computational expensive bootstrapping process that makes FHE impractical to most control systems.
Kim et al.~\cite{kim2016encrypting} propose FHE to implement a dynamic output feedback controller using multiple controllers to avoid the bootstapping delay. However, another inherent issue with FHE is that the ciphertext expansion might be up to $10000:1$ for an acceptable level of security of $100$ bits~\cite{chillotti2020tfhe}.
% the authors of \cite{kim2016encrypting} employed fully-homomorphic encryption to enhance the security of control system with a special focus on conventional controller. 
%
Pailier's homomorphic encryption (PHE, supporting only encrypted additions) has also been proposed to implement a variety of controllers~\cite{tran2020implementing, murguia2020secure}. However, due to memory issues related to the state of dynamic encrypted controller (i.e., the number of bits required for its representation grows linearly with the number of iterations), the solution is limited to the use of %\mm{the rest of this sentence is unclear} 
resetting dynamics control laws. 
PHE was also proposed by Lin et al.~\cite{lin2018secure}, where it is shown that only a subset of the 
%\mm{what is real control gain?} 
real control gains can be used  (according to the available bits).

Overall, existing solutions pose several limitations in terms of security/privacy/deployability to networked control systems. Moreover, no solutions have been proposed to protect CPSs against a malicious cloud operator, or malware that might be able to compromise the integrity of the control algorithm running on the cloud server.

\section{Our Proposal}
\begin{figure}[htb]
\centering
        \includegraphics[width=0.7\columnwidth
        ]{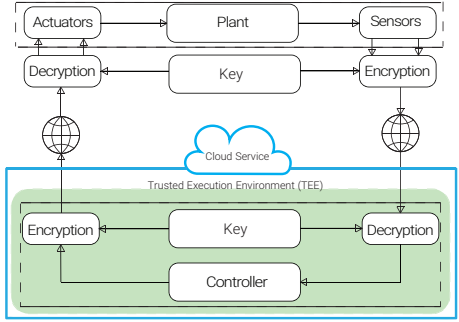}
        \vspace{-0.3cm}
        \caption{Proposed solution based on a trusted execution environment.}
        \label{Fig3}
\end{figure}

\vspace{-0.3cm}
\label{ProposedModel}
%In what follows, we introduce a novel architecture capable of overcoming such drawbacks. 
The objectives of our proposal are: 
secure the cloud-based CPSs against all the cyber-threat discussed in Section~\ref{sec: ncss}, and  reduce the impact on the design and implementation of existing control strategies. The proposed secure control architecture has two essential components (see Fig.~\ref{Fig3}): an authenticated encryption scheme for securing the communication channels, 
%between the plant and the cloud, 
and a TEE where the control logic is executed and the secret cryptographic keys, used by the authenticated encryption scheme, are stored.

First, we resort to authenticated encryption schemes (cf.~\cite{patel2009improving}) to ensure the integrity and the confidentiality of the control signal and sensor measurements exchanged between the plant and the controller. The used encryption scheme must be characterized by an inherent latency much smaller than the control-loop sampling time. The latter requirement is essential to ensure that the encryption scheme does not affect the control-loop system's stability. Second, a trusted execution environment (TEE) is used to protect the controller's operations in the cloud service. Generally speaking, a TEE refers to a hardware-based solution capable of ensuring that no malicious cloud entities (e.g., malware or a malicious cloud operator) could interfere with the execution of the control algorithm or with the memory associated with it.  Moreover, if encryption/decryption operations are executed inside the TEE, where the keys are also protected by the TEE, then a malicious cloud administrator also cannot access the keys.
TEE may also provides some other advantages  such as measuring the integrity of the launched processes, measuring the origin of the TEE and current state of the TEE (attestability), and recovering the state of the TEE to a known good state after any corruption (recoverability). The presence of a TEE on the plant side is not  required for our threat model. However, it is desirable in a scenario where the local computing platform (e.g., SCADA system) could be subject to cyber-attacks. Several solutions have been proposed in the literature (not in CPS) %\mm{are these examples related to CPS? if so, mention it}
%\todo[inline]{W: these have not been used in CPS. Maybe we can mention in which specific literature they belong to? (security, computing, not sure which one)}
using different TEE implementations, e.g., Intel SGX \cite{costan2016intel}, ARM TrustZone \cite{infrastructure2004technical}, AMD SEV \cite{kaplan2016amd}, Hardware Security Module (HSM) \cite{varia2014overview}, and secure co-processors \cite{bajaj2013trusteddb}. Although all these solutions provide strong security mechanisms, not all can be used in our design (e.g., HSMs do not support remote attestation as opposed to Intel SGX).  %For example, if HSM is used to implement the control logic, then, given a computed control input, it is not possible to  provide a certification/proof that such a result is the outcome of the agreed upon control logic. On the other hand, if Intel SGX is used, we can obtain cryptography evidence  that the controller output is consistent with the intended control algorithm. 

\begin{figure}[htb]
\centering
        \includegraphics[width=1\columnwidth]{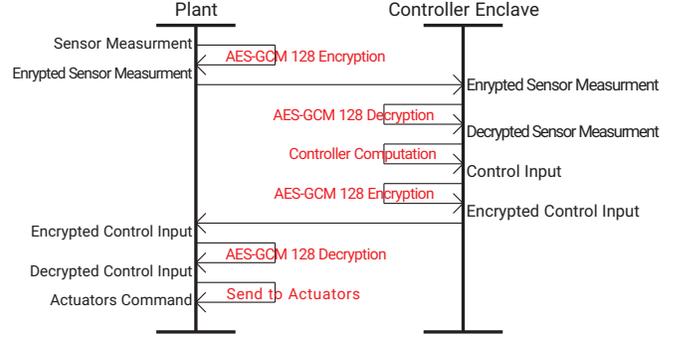}
        \caption{Data Flow in the Proposed Solution.}
        \vspace{-0.3cm}
        \label{Fig:Work_flow}
\end{figure}

% \todo[inline, color=green]{Is there any issue/potential issue common to any TEE (not only to SGX)? If yes you can mention them here briefly.}

% \todo[inline, color=orange]{One of the most important problem with Intel SGX is limitation of accessible memory. So, if the controllers data and equation become too large, it can be problem that how to keep all of them in an enclave. Other problem that comes to my mind is the vulnerability of SGX against some kinds of attacks like channel-side attack and cache attack. I do not know this attack too much, but I have seen many papers about this attacks on SGX in "Scholar". But, here, I think we can assume that these attacks can not be applied on SGX. I mean we can assume that SGX is fully secured.}

% \todo[inline, color=green]{you can provide more details on how SGX works. You can have a flow diagram to show what is happening with the presence of SGX. SGX is the core of the proposed implementation and we need to provide way more details.\\
% Use as "template model" this paper\\
% https://arxiv.org/pdf/1810.01651.pdf\\
% In the paper there are some security analysis that you can also do here
% }
% \todo[inline, color=orange]{Resolved.}
%********************************************************************
\section{Implementation}
\vspace{-0.3cm}
%\mm{There is nothing about implementation here - just what environment/parameters you choose. Also, how/when do you verify the control logic's integrity? How the keys are provisioned? Operational steps are not at all clear from the figures or text in Sec 4 and 5. This must be improved. /{\color{red} Amir: Resolved.}}
% \todo[inline]{W:Amr please checks Amir's answer. According to him, Mannan's comment should not be addressed. Do you agree? Also we dont have space to comprehensively describe Fig. 3}
We use Intel SGX as TEE for its capability of providing a cryptographic attestation to ensure the integrity of the execution of the controller algorithm, even in the presence of a malicious cloud admin or a compromised cloud operating systems (e.g., by a malware). %Intel SGX is a hardware-based security solution that utilizes encryption to change the way memory is accessed \cite{costan2016intel}. 
To keep code and data secure, SGX provides an isolated execution environment, and encrypted memory. This secure container is called an ``enclave'' and everything else outside the enclave is assumed to be insecure. 
Two main functions are available to interact with the enclave, namely Encalve Call (E-Call) and Out Call (O-call). E-Call is used to call, from outside the enclave, a function implemented inside the enclave. On the other hand, O-call is used to call, from inside the enclave, a function implemented outside the enclave. 

For the implementation of the authenticated encryption, AES Galois/Counter Mode (GCM) is used. This algorithm is a good candidate for CPSs
% \Amir{ I think it is better to use the term "CPS" here, if we want to use the term "CPS" at the begining of the paper where Dr. Mannan left a comment.} 
because of its high throughput and low latency~\cite{koteshwara2017fpga, Arun2015ImplementationOA}. 
%An initialization vector (IV) is used in AES-GCM as a nonce. This IV should be different in each encryption operation. The inadvertent reuse of the same nonce with the same key, undermines data confidentiality.
First, we need to create an enclave and allocate memory for the Enclave Page Cache (EPC). The process starts with the attestation of both the enclave (validity of the CPU's SGX support) and the code (validity of the binary executed within the enclave as the controller logic). During the attestation, entities also establish a secure session key. After these initialization operations, data transmission will be started between the participating entities, encrypted under the session key.
The data flow for a single control loop is shown in Fig.~\ref{Fig:Work_flow}. In particular, the sensor measurements are encrypted on the plant side. Then, these encrypted sensor measurements are sent to the cloud over  the communication channel. The authenticity of the received measurement is checked inside the enclave, where then the controller logic is also applied to the  decrypted measurements. 
% In the controller side and inside the cloud service, encrypted sensor measurements, are decrypted inside the controller enclave.
The evaluated controller output is then encrypted (inside the enclave) before it is sent to the actuator through the communication channel.
% To send the control input of the system calculated in the controller back to the plant, to be a command for actuators, it has to be encrypted again. Hence, the control input is encrypted inside the controller enclave and is sent back to the plant side through the communication channel. 
Finally, the encrypted control input is decrypted by the actuator and applied to the  plant.  
% to be decrypted, so it can be used by the actuators. After the control input is decrypted, it becomes a command for actuators. 

% \todo[inline, color=green]{ The security analysis applies to general trusted computing platforms or only to SGX? If only to SGX then you need to move this analysis after you discuss SGX. \\
% Moreover, to show that we are aware of side channel attacks against SGX, you can mention a sentence like this:  "We are aware of recent side channel attacks on Intel SGX. Defending against these attacks is outside the scope of this paper".\\
% https://software.intel.com/content/www/us/en/develop/articles/intel-sgx-and-side-channels.html
% }
% \todo[inline, color=orange]{ I think this is not the best place in the paper for security analysis, however I think there is just one other choice to add it as a subsection in section 6 and before the simulation subsection, but then I believe the title of section 6 has to be changed/}
% \todo[inline, color=green]{Move the security analysis after section 4. After section 4, you can have a section named something like ``Implementation through SGX and Security Analysis" where you first discuss why we want to use SGX (what we now have at the beginning of section 5) and then you do the security analysis. After this section, you can have a section called Simulation Results (only about the simulation)
% }
% \todo[inline, color=orange]{Resolved.}

\section{Security and Performance Evaluation}
\vspace{-0.3cm}
%\subsection{Security Analysis}
We now discuss the security properties of the proposed solution. (i)  
\textit{Confidentiality:}
Data sent through the communication channels are encrypted with AES-GCM. Therefore, network eavesdroppers are unable to decrypt the transmitted control signals and sensor measurements. Moreover, control operations and encryption/decryption operations are performed within the enclave, avoiding the possibility that a malware or cloud administrator could intercept the plaintext signals or acquire the keys. (ii) \textit{Integrity:}
By exploiting the message authentication code (MAC) tag in AES-GCM, it is possible to verify the integrity of the transmitted data (i.e., detect if an attacker has manipulated the transmitted data). Another aspect of integrity is to make sure that the controller logic is not manipulated by the cloud provider before the code is executed within the enclave. For this purpose, an attestation operation is performed to make sure that the code executed in the enclave is exactly that is sent to the cloud service by the system admin. To improve code obfuscation (i.e., hiding the control logic from the cloud operator), the proposed solution in~\cite{bauman2018sgxelide} can be used. Note that the controller's runtime state remains always protected by SGX's memory encryption.
Moreover, since the controller is executed inside SGX, the integrity of the control algorithm is also ensured. (iii)
\textit{Authentication:}
The remote attestation feature of Intel SGX is used on the plant side to establish a secure and authenticated communication channel with the enclave in the cloud and ensure that the remote enclave is trusted. The MAC tags also 
% \mm{Why "can be"? This is not an option./{\color{red} Amir: I have changed it to "is used."}} 
is used by both entities (plant and controller) to make sure that the received messages are obtained by a trusted entity. (iv)
\textit{Freshness:}
The uniqueness of the AES-GCM IV is used to guarantee freshness of each message.
Defending against side-channel attacks against Intel SGX~\cite{brasser2017software} is outside the scope of this paper. In the case of necessity of storing data by the controller (depend on the controller logic),
%on the cloud, 
to mitigate rollback attacks on the sealed data, Monotonic Counter (MC) of Intel SGX can be used to guarantee that the sealed data is the latest copy.
% \mm{Are there saved states at the cloud-end? Can they be replayed?} \Amir{I have not saved any specific number in the controller for the state of the system. The only numbers that are kept inside the controller (Which is implemented inside the SGX) is the control command and previous estimation of the states of the system. I think that they can not be replayed because: 1- the estimations can not be replayed because they are not sent out of the controller, so no one has access to them to collect them and reply them. 2- the control command can not be replayed too, because the controller keeps it inside the SGX and it is not a input for controller.}
% \todo[inline]{W:Amr please double-check if Amir's answer is right. I am not sure if Mannan means if there are variable saved in the cloud outside the enclave?}

%\vspace{-0.25cm}
\subsection{Performance Evaluation}
\vspace{-0.2cm}
%\section{Simulation Results}
%\label{sec:simul}
%In this section, we evaluate the performance of the proposed architecture by means of a simulation example. 
\subhead{System setup}
As a testbed, we use the Quadruple Tank Process (QTP) system from Johansson~\cite{johansson2000quadruple}, which is often used as a benchmark for control systems applications.
% \mm{No justification is given for this choice /{\color{red} Amir: I beleive that is a popular test bed used in many papers. also it is a good simple example of a small industrial control system or CPS.}}.
The system consists of four water tanks where $h_i, i \in 1,2,3,4 $ represents the level of water in each tank and also represents the states $x$ of the system, i.e., $x=[h_1,\,h_2,\,h_3,\,h_4]^2\in \rr^4$. There are two sensors that measure the level of water inside tanks 1 and 2, i.e., the output measurement vector is $y=[0.5h_1,\,0.5h_2]^T\in \rr^2$. Moreover, the system is equipped with two pumps and the applied voltage $v_1, v_2$ are the inputs $u$ of the system, i.e., $u=[v_1,\,v_2]^T$. 
We have linearized the system model 
% \mm{by who?/{\color{red} Amir: linearization of the system equations is done in the cited paper. (Johansson) }} 
around the equilibrium pair ($x_{eq}=[12.4,\,12.7\,,1.8,\,1.4]^T$  $u_{eq}=[3,\,3]^T$) and discretized it using a sampling time $T_{s}=0.1\sec$. The linearized model $x(k+1)=Ax(k)+Bu(k),$ $y(k)=Cx(k)$ and its matrices $A,B,C$ can be easily obtained following~\cite{johansson2000quadruple}. 
% expressed as follows
% %ww
% \begin{align}
%   & x(k+1)=Ax(k)+Bu(k)+v(k) \\ 
%   & y(k)=Cx(k)+w(k)
%   \label{eq:state_equation}
% \end{align}
% %
% where ${{x}_{k}}\in{{\mathbb{R}}^{n}}$ is state vector, ${{u}_{k}}\in {{\mathbb{R}}^{m}}$ the control input vector ,${{y}_{k}}\in {{\mathbb{R}}^{p}}$ the output vector and $v(k)$ and $w(k)$ are independent and identically distributed process and measurement noise with Gaussian distribution. The state-space matrices are here omitted because they can be obtained from \cite{johansson2000quadruple}.
%
% \begin{align}
% &A=\left[ \begin{matrix}
%   0.9984 & 0 & 0.0042 & 0  \\
%   0 & 0.9989 & 0 & 0.0033  \\
%   0 & 0 & 0.9958 & 0  \\
%   0 & 0 & 0 & 0.9967  \\
% \end{matrix} \right]\\&
%  B=\left[ \begin{matrix}
%   0.0083 & 9.9969\times{{10}^{-6}}  \\
%   5.1966 \times {{10}^{-6}} & 0.0063  \\
%   0 & 0.0048  \\
%   0.0031 & 0  \\
% \end{matrix} \right]
% \\&
%  C=\left[ \begin{matrix}
%   0.5 & 0 & 0 & 0  \\
%   0 & 0.5 & 0 & 0  \\
% \end{matrix} \right]
% \label{eq:state_space_matrices}
% \end{align}
%------------------------------------------------------------------------%
The plant is regulated by means of dynamic output feedback controller consisting of a 
a Luenberger Observer and an optimal Linear Quadratic (LQ) controller. The state-estimator operations are described by the discrete-time system
$
\hat{x}(k+1)=A\hat{x}(k)+Bu(k)+L(y(k)-C\hat{x}(k))
$
where $\hat{x}(k)$ is the estimation of the state $x(k)$ and the correction gain  is given by 
$
L=\left[ \begin{matrix}
   0.78 & 0 & 0.32 & 0  \\
   0 & 0.78 & 0 & 0.32
\end{matrix} \right]^T.
$
%
% $
% L=\left[ \begin{matrix}
%   0.7815 & 0  \\
%   0 & 0.7816  \\
%   0.3190 & 0  \\
%   0 & 0.3199  \\
% \end{matrix} \right].
% $
%
The LQ controller logic is computed as $u=K(x-{{x}_{eq}})+{{u}_{eq}}$ where the stabilizing gain is given by
$
K=\left[ \begin{matrix}
   27.547 & -0.054 & 0.468 & 0.086  \\
   0.023 & 28.441 & 0.143 & 0.507  \\
\end{matrix} \right].
$

\noindent The dynamic output feedback controller operations have been implemented by utilizing an Intel SGX running on an Intel Core i7-6700 CPU, 3.40GHz, with 4 cores and 8 threads and 16 GB of RAM, using 64-bit Windows 7.%
\subhead{Measurements}
We have conducted a series of  measurements to evaluate the computation times required by different components of the proposed solution (see the data flow in Fig.~\ref{Fig:Work_flow}). 
% \mm{Is it true?/{\color{red} Amir: I believe that it is not proper to use the term "impossibility." regarding this exact phrase of the cited paper: "The  current  generation  of  SGX  does  not  support  the  use of  the  RDTSC instruction or any other native timing facilities inside enclaves. Intel has later released a microcode update to counter this problem, allowing for the RTDSC instruction to execute inside enclaves. We are however unsuccessful, at the time of writing, in obtaining a firmware update specific to our SKU through the correct OEM." But I believe that the method that we used for our measurements is better for our paper. Because we are measuring the total overload of the intel SGX. So the in/out time is important for us, because is takes time in each operation loop and the time is significant and it is not negligible. Hence, it has to be measured.} } 
% \todo[inline]{W:Amr maybe it is sufficient to say that for the avaialbe SGX setup, direct CPU time measurements are not possible? (see text in blue below)}
The reported CPU measurements have been obtained using the approach  proposed in~\cite[Fig. 1]{gjerdrum2017performance}, i.e., an O-call function is used as a stopwatch. As a result, the time measurements  in Table~\ref{table2:time_complexity} include an extra time representing the CPU time required to return to the enclave from an O-call and exit from it. We denote this time by $\Delta t$. To mitigate the presence of $\Delta t$ in the measurements, we repeated each operation inside the enclave 1000 times and then calculate the average. $\Delta t$ is also measured separately. The numerical results show that the two dominant factors are $\Delta t$ and the control algorithm CPU time. Indeed, the average total CPU time required by both the secure and insecure implementations are around
%$941.9419\mu s$  and $478.4785\mu s,$ 
$905 \mu s$  and $479\mu s,$ respectively. The obtained results  confirm that the computational overhead introduced by the use of Intel SGX does not affect the feasibility of the control strategy. Moreover, given that the introduced overhead is in the milliseconds' range, the proposed SGX-based secure architecture is believed to be affordable for a large class of  cloud-based control systems applications.

\begin{table}[h!]
\centering
 \begin{tabular}{||c c||} 
 \hline
 Operation & Time ($\mu s$) \\ [0.5ex] 
 \hline\hline
 Enclave creation & 8368.4 \\ 
 Dynamic output feedback controller  & 466.7  \\
 AES-GCM encryption & 1.8 \\
 AES-GCM decryption & 1.4 \\
 $\Delta t$ & 435.4  \\ % [1ex] 
 \hline
 \end{tabular}
\caption{Average time for different operations of the SGX-based solution}
\vspace{-0.25cm}
\label{table2:time_complexity}
\end{table}
% \begin{table}[h!]
% \centering
%  \begin{tabular}{||c c||} 
%  \hline
%  Operation & Time ($\mu s$) \\ [0.5ex] 
%  \hline\hline
%  Enclave creation & 8368.4 \\ 
%  Dynamic output feedback controller  & 466.67  \\
%  AES-GCM encryption & 1.7968 \\
%  AES-GCM decryption & 1.3854 \\
%  $\Delta t$ & 435.435  \\ [1ex] 
%  \hline
%  \end{tabular}
% \caption{Average time for different operations of the SGX-based solution}
% \label{table2:time_complexity}
% \end{table}
%***************************************************************************************%
\vspace{-0.2cm}
\section{Conclusion}
\vspace{-0.3cm}
\label{sec:cnc}
We  proposed a solution to secure cloud-hosted CPSs. In particular by resorting to authenticated encryption and  a trusted execution environment, we showed that the proposed networked control scheme is secure again different attacks against its security and privacy. We verified the effectiveness of such a scheme by means of numerical simulations obtained considering Intel SGX, where we performed different benchmarks to evaluate the computational burden associated to the trusted control scheme implementation. The obtained results show good promise in terms of real-time performance and simplicity of implementation in CPSs applications.
%, encouraging to further develop such a security solution.

% shown an implementation of such a scheme by considering Intel SGX and a standar
% The model proposed in this paper, in addition to eliminating drawbacks in the previously proposed methods, provides system security and privacy at a higher level. In this model, the attacker is not able to eavesdrop messages, integrity of messages is guaranteed, and the controller of the system and keys are not accessible to the attacker. Another important concern in addition to security is the efficiency of the model. The extra load imposed on the system should not affect the performance. The simulation results show that the proposed model is optimal enough to be implemented even on processors with not very high computing power.
%************************************************************************************%

\vfill
\pagebreak
% References should be produced using the bibtex program from suitable
% BiBTeX files (here: strings, refs, manuals). The IEEEbib.bst bibliography
% style file from IEEE produces unsorted bibliography list.
% -------------------------------------------------------------------------
\bibliographystyle{IEEEbib}
\bibliography{strings,refs}
%\printbibliography
\end{document}